\documentclass[final,3p,times]{elsarticle}

\usepackage{graphicx}
\usepackage{amssymb}
\usepackage{lineno}
\usepackage[colorlinks=true,urlcolor=blue]{hyperref}

\begin{document}

\begin{frontmatter}


\title{Citation network centrality: a scientific awards predictor?}
\author{Osame Kinouchi*, Adriano J. Holanda and George C. Cardoso}

\address{Departamento de F\'{i}sica, FFCLRP, Universidade de S\~ao Paulo\\ Ribeir\~ao Preto - SP, 14040-901, Brazil}

\begin{abstract}
  The $K$-index is an easily computable centrality index in complex
  networks, such as a scientific citations network. A researcher has a
  $K$-index equal to $K$ if he or she is cited by $K$ articles that have
  at least $K$ citations.  The $K$-index has several advantages over
  Hirsh's $h$-index and, in previous studies, has shown better
  correlation with Nobel prizes than any other index given by the {\em
    Web of Science},  including the $h$-index.  It is plausible that researchers who are the most connected to other scientifically well-connected researchers are the  most likely to be doing important work and more likely to be awarded major prizes in a given area. However, the correlation found does not imply  causation.
  Here we perform an experiment using the $K$-index, producing a shortlist of twelve candidates for major scientific prizes, including the
 Physics Nobel award, in the near future. For example, our
  top-12 $K$-index list naturally selects the 2019 Nobel laureate James Peebles. The list
  can be updated annually and should be compared to 
  laureates  of the following years.
\end{abstract}

\begin{keyword}
Complex networks\sep
Node centrality\sep
Hirsch index\sep
K-index \sep
Nobel prizes.\\
$^*$ Corresponding author: okinouchi@gmail.com
\end{keyword}

\end{frontmatter}

\newpage

\section*{Highlights}
\begin{itemize}
\item We propose that the $K$-index could be a good network centrality index for the physics community and relevant to predict the likelihood of scientific prizes;
\item We propose an experiment where a list of highly cited candidates is refined to predict Physics Nobel Prizes in the near future;
\item We present a list with twelve candidates with highest $K$-index mostly based on an initial list of $133$ physicists from {\em Clarivate Highly Cited Researchers 2019} (HCR). 
\item We find that the 2019 Nobel laureate James Peebles is
ranked  $11th$ on a list where 
graphene researchers, of an area already awarded in 2010, are filtered out.
\item We present the $K$ versus $h$ plane for the HCR list.
\end{itemize}

\section{Introduction}
\label{S:1}

Statistical physicists have made important contributions to the
interdisciplinary area of complex
networks~\cite{newmannetworks,barabasi2016network}.  In particular,
physicists have intensively studied scientometric networks thanks to the availability of large and reliable data
banks~\cite{newman2001structure,barabasi2002evolution,wang2008measuring,ren2012modeling,clough2016dimension,xie2016geometric}.
Indeed, an important advancement for the area came with  the introduction of the
$h$-index by physicist Jorge E. Hirsch~\cite{hirsch2005index}.  A
researcher has $h$-index $h$ if he/she has published $h$ 
papers each one with at least $h$ citations.  
Centrality indexes proposals for
citation networks experienced a boom after
the introduction of the $h$-index~\cite{batista2006possible, egghe2006improvement,hirsch2007does,
  schreiber2010twenty,todeschini2016handbook}.

A decisive advantage of the $h$-index over its competitors is its
ease of calculation. However, it also is known that the $h$-index has
several drawbacks. For example, if a researcher has published a small
or moderate number $N$ of papers, then necessarily $h \leq N$, even if
every paper is of very high quality and has received thousands of
citations.

Recently, we have proposed the $K$-index,  a centrality index that is complementary to $h$-index and also very easy to calculate 
in the {\em Web of Science} ({\em
WoS})~\cite{kinouchi2018,kinouchi2018k}.  In these publications, we
verified that Physics and Physiology Nobel Prizes laureates have very
high $K$-index (often above $K=300$) but sometimes moderate
$h$-index. It is very plausible that researchers who are the most scientifically connected to other scientifically well-connected researchers are the  most likely to be doing very relevant research and are more likely to be awarded major prizes in a given area. However, this is only a correlation, and the growth of the
$K$-index could have occurred after the acceptance of the prize.

Here we propose to test  $K$-index's  predictive power by using
it to refine the {\em Clarivate Highly Cited Researchers
2019} (HCR) list of candidates to the 2019 Physics Nobel Prizes.
Our task is a hard one since it depends only on a brute 
correlation between a scientometric index and the awards, and does not
take into account nuanced and sociological guesses about the actual 
candidates for scientific prizes, including the Nobel prize.

\section{Materials and Methods}

\subsection{The Highly Cited Researchers list of
Clarivate Analytics}
\label{SS:2.1}

As a primary source, we used {\em Clarivate Highly Cited Researchers
  2019} (HCR) list to furnish an initial sample of $133$ candidates
that have Research ID or Orcid.  The methodology used by HCR to
achieve this sample list is not of our concern now, and can be found in:

https://hcr.clarivate.com/methodology/purpose-and-methodology/ 

To the HCR list we added the three 2019 Physics Nobel Prize laureates.
Our data and automated ranking script is available for public use at:

https://github.com/ajholanda/k-index/blob/master/disclosure.md

Our task is to refine the HCR list by using the $K$-index.  We will
produce a shortlist of twelve candidates which is the maximum number
of Nobel laureates for a period of four years and about $8.8\%$ of the
original HCR {\em 2019} list.

For comparison, the $K$ and $h$-indexes
for the $136$ physicists from the HCR list are presented in Fig.~\ref{fig1}. The $h$-index is
furnished directly by the {\em WoS}.

\subsection{Calculation of the $K$-index}
\label{SS:2.3}

The $K$-index has been devised to measure the
impact of the papers that cite a researcher,  not just to measure the quantity or distribution of citations.  If a maximum number of
$K$ papers cite a given author, each one with at least $K$ citations, 
then the researcher has $K$-index equal to $K$~\cite{kinouchi2018,kinouchi2018k}.

Centrality indexes that tried to improve the $h$-index, in general,
involve impractical calculations~\cite{todeschini2016handbook}.  The
decisive advantage of the $K$-index is that it is easily determined by
simple inspection of the \emph{WoS} platform. We presume that other
platforms like Google Scholar Citations could also be easily adapted
to provide $K$ automatically.

On the \emph{WoS}, currently, one can obtain the $K$-index 
of a researcher from the following simple steps:
\begin{itemize}
\item Search the production of a given author;
\item Click on the link \emph{Create Citation Report};
\item Click on the link \emph{Citing Articles} ($CA$)  (or {\em Citing
  Articles without self-citations}, if desired);
\item Have the list of citing articles ranked from the most cited
  (defined as rank $r = 1$) to the least cited (that is the default
  ranking presented by \emph{WoS});
\item Compare the article rank $r$ (on the left) with
its citation count $c(r)$ on the right. When $r \leq c(r)$
but $r+1 > c(r+1)$, stop: the $K$-index is $K=r$.
\end{itemize}

\section{Results and Discussion}

\begin{figure}[ht]
\begin{center}
\includegraphics[width=0.9\linewidth]{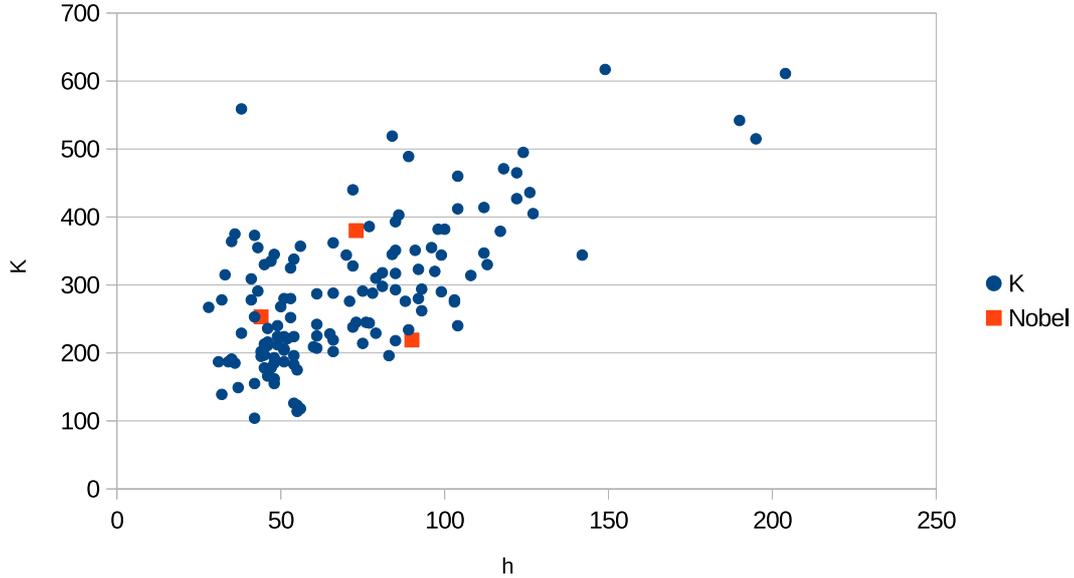}
\end{center}
\caption{ $K$-index vs $h$-index for the HCR list (blue circles)
and 2019 Physics Nobel Prize laureates (red squares).}
\label{fig1}
\end{figure}

In table I, we present twelve candidates from 
the {\em Clarivate} HCR list
ranked by the $K$-index.  In Fig.~\ref{fig1}, we present the $K$
versus $h$ plane for the $133$ researchers from the HCR list (blue
circles) and the 2019 Nobelists (red squares). 
We see that the laureates have $K$ and $h$-indexes comparable
to the HCR group.

Our objective with Fig.~\ref{fig1} is to show that scientists 
with high $K$ do not necessarily have high
$h$ and vice-versa. $K$ and $h$ have complementary information.
Fig.~\ref{fig1}  should  be compared to the $K$ vs. $h$ plots
in~\cite{kinouchi2018,kinouchi2018k}, where $28$
Nobel laureates  show  $K$ values well above other scientists' 
of similar $h$. However, it is not clear how much their $K$ indexes
have grown after receiving their prizes. A considerable inertial growth is a plausible hypothesis and correlation effects are difficult to
separate.

\begin{table}[ht]
\centering
\begin{tabular}{c l c c}
\hline
\textbf{Rank}   &\textbf{Name} & \textbf{K} & \textbf{h}\\ \hline
 1 & Paul Alivisatos         & 617    & 149 \\
 2 & Michael Graetzel        & 611    & 204 \\ 
 3 & Sergey V Morozov        & 559    &  38 \\
 4 & Younan Xia              & 542    & 190 \\
 5 & Philip Kim              & 519    & 84 \\
 6 & Zhong Lin (Z.L.) Wang   & 515    & 195 \\
 7 & Yi Cui                  & 495    & 124 \\
 8 & Mikhail I. Katsnelson   & 489    & 89 \\
 9 & Yang Yang               & 471    & 118 \\
10 & Phaedon Avouris         & 465    & 122 \\
11 & Alex K. Zettl           & 460    & 104 \\
12 & James Hone              & 440    & 72  \\
\hline\end{tabular}
\caption{List of twelve Nobel Prize candidates as ranked by the $K$-index}
\end{table}

Fig.~\ref{fig1} intends not to be correlational but
predictive. All candidates have high citation rates and comparable
$h$-index. However, we have chosen the top twelve $K$ as our (crude)
test of the predictive power of the $K$-index.

A limitation of our study is that our original sample is from {\em
Clarivate} HCR: if possible nominee candidates are not
initially on the list, then our present $K$ ranking 
cannot detect and select them. This indeed occurred in the
2019 Nobel Prize: the laureates P. James E. Peeble, Michel Mayor
and Didier Queloz were not included in the 2019 HCR list.
Here, we added them to the list and compared their $K$ and $h$-indexes
to the other researchers. Of course, it is possible that other
high $K$ researchers exist but have not been included in the HCR
list, and their inclusion would change the ranking by $K$
(from here, the $K$-rank). 

Another limitation of our study is that the chosen names 
are, in principle,
uncorrelated, that is, do not refer to a common discovery or 
a similar research topic. By contrast, the Nobel Prize for a given year
is typically awarded to researchers who have made 
progress on similar topics or discoveries.  
Also, it is possible that researchers with very 
high $K$ or $h$ will not be
laureate because they work on a topic that has 
already been awarded in the recent past.  

Our methodology can be adjusted
to incorporate this kind of information. For example,
we have noticed a very high proportion of authors 
with large $K$ in the area
of materials science, especially graphene research -- which
has already been an awarded topic in 2010. 
To correct for this bias, 
we have produced a second list where graphene researchers are removed.
The new list with twelve candidates and with graphene scientists filtered out is given in Table~2 and has seven 
new names as a replacement.

\begin{table}[ht]
\centering
\begin{tabular}{c l c c}
\hline
\textbf{Rank}   &\textbf{Name} & \textbf{K} & \textbf{h}\\ \hline
 1 & Paul Alivisatos          & 617    & 149 \\
 2 & Michael Graetzel         & 611   & 204 \\
 3 &  Younan Xia              & 542    & 190 \\
 4 & Zhong Lin (Z.L.) Wang    & 515    & 195 \\
 5 & Yang Yang                & 471    & 118 \\
 6 & Mohammad K Nazeeruddin   & 436    & 126 \\
 7 & Naomi Halas              & 427    & 122  \\
 8 & Zhenan Bao               & 412    & 104  \\
 9 & John Rogers              & 405    & 127 \\
10 & Arthur J. Nozik          & 386    & 77 \\
11 & {\bf P. James E. Peebles} & 380    & 73 \\
12 & Peter Zoller             & 379   & 117  \\
\hline\end{tabular}
\caption{List of twelve Nobel Prize candidates as ranked
by the $K$-index with graphene scientists filtered out. 2019
laureate James Peebles has  rank $11^o$  when compared to
researchers that pertain to the HCR list.}
\end{table}

We conclude that, if the 2019 Nobel laureates were included
in the HCR list, James Peeble (rank $11^o$ of
Table~2) would pertain to our
top twelve filtered list.  So, in a sense, we
can say that our very crude method 
(ranking by the $K$-index plus a filter to already awarded
topics) predicted one of the 2019 Nobel laureates.

In the unfiltered list, we have that James Peebles ($K=380$ and $h=73$) has
$K$-rank of $23^o$ and $h$-rank of $53^o$,  Michel Mayor ($K = 253$ and $h=44$) has
$K$-rank of $76^o$ and $h$-rank of $86^o$ and 
Didier Queloz ($K=219$ and $h=90$) has $K$-rank of $96^o$ and $h$-rank of $42^o$. The low $K$-rank
of Mayor and Queloz is understanable: the area of
exoplanets research is small and citations cannot
be accumulated.
We conclude that
ranking by the $K$-index has outperformed ranking
by the $h$-index in the case of James Peebles and Michel Mayor,
but has not outperformed $h$-ranking for Didier Queloz.

The average and standard deviation 
values for $K$ and $h$ in the list are
$K = 287 \pm 104$ and $h = 71 \pm 33$. 
So, although 
the 2019 laureates have not been included in the
original HCR list due to the (defective?) 
Clarivate's 
methodology, their $K$-index and
$h$-index have the same level
of the other researchers and
stay at less than one 
standard deviation of the HCR mean.
We notice also that the coefficient of variation
for the $K$ index is $CV_K = \sigma_K/\bar{K} =
0.36$ and for the $h$-index is $CV_h = \sigma_h/\bar{h} = 0.46$. This means that the
$K$-index characterizes and defines better the
HCR sample than the $h$-index.

\section{Conclusion}
\label{S:4}

It is an open question whether bibliometric 
information can have predictive power
for scientific prizes. Prizes  denote qualitative scientific
recognition at the sociological level, where human
factors are very important. Nobody would think that a prize should be decided by ranking the production of scientists by some automatic metric. At the same time,
prizes intend to recognize original contributions whose impact is
reflected in the bibliometric indexes, so it is plausible that
predictive information is hidden in these indexes.

From a list of highly cited researchers, we proposed 
twelve candidates
for the 2019 or following years Physics Nobel Prizes. 
We have presented a naive ranking and also
an improved ranking where a citation bias for materials scientists
studying graphene was filtered out.
In this new list, the 2019 Physics Nobel laureate James Peeble
appears having the $11^o$ rank.
Our list of candidates can be
updated and also used in future years.

The fact that the 2019
laureates were not included to the original 
HCR list means that Clarivate's ranking methodology has problems:
for example, the HCR is based only in highly
cited papers in the 2006-2016 period but the
seminal papers of the 2019 laureates are from
the 90's or earlier. For an ideal test,
we would need a complete database of $K$-indexes
for all researchers with ResearchID or Orcid.
In any case, we have showed that
the Nobel laureates have $K$ and $h$
comparable (within one
standard deviation from the average) 
to the rest of the HCR list.

The shortlisting  study of this paper 
could be extended to other
scientific prizes and other scientific
disciplines. The only difference is that the sample of initial
candidates should be selected in accord with the 
specific scientific area. 
These
predictive tests, perhaps in the form of annual contests,
could be useful benchmarks for evaluation of
centrality indexes that can then be used in other, less monitored and less well-studied, complex networks.

\section*{Acknowledgments}
O.~K.~acknowledges financial support from the National Council for Scientific and Technological Development (CNPq) and CNAIPS~--~Center
for Natural and Artificial Information Processing Systems (University of Sao Paulo). This
paper results from research activity performed at The S\~ao Paulo Research Foundation (FAPESP) Center for Neuromathematics (FAPESP grant 2013/07699-0).

\section*{Declarations of interest} 
None.

\bibliographystyle{apalike}
\bibliography{knobel}

\end{document}